\documentclass[epj,referee]{svjour}
\usepackage{amsmath,amssymb,epsfig,subfigure}
\begin{document}
\title{Simultaneous and Sequential Synchronisation in Arrays}
\author{G. Ambika\thanks{\emph{ambika@iucaa.ernet.in}} \and K. Ambika\thanks{\emph{ambikak2002@yahoo.com}}
}
\institute{Department of Physics, Maharaja's College, Ernakulam, Kerala, India.}
\abstract{
We discuss the possibility of simultaneous and sequential synchronisation in
vertical and horizontal arrays of unidirectionally coupled discrete systems.
This is realized for the specific case of two dimensional Gumowski-Mira maps.
The synchronised state can be periodic, thereby 
bringing in control of chaos, or
chaotic for carefully chosen parameters of the participating units. The 
synchronised chaotic state is further characterised using variation of the 
time of synchronisation with coupling coefficient, size of the array etc.
In the case of the horizontal array, the total time of synchronisation can be
controlled by increasing the coupling coefficient step wise in small bunch of 
units.
}  
\PACS{
{05.45.Xt}{Synchronization; coupled oscillators} \and
{05.45.-a}{Nonlinear dynamics and nonlinear dynamical systems}
}
\maketitle

\section{Introduction}
Synchronisation of the dynamical variables of coupled systems is an important
nonlinear phenomenon where intense research is being concentrated
recently \cite{sesy-ref1}. This is probably because of its engineering 
applications like spread spectrum and secure data transmission using chaotic
signals \cite{sesy-ref1,sesy-ref2,sesy-ref3}, control of microwave electronic
devices \cite{sesy-ref4}, graph colouring etc. Also communication between
different regions of the brain depends heavily on the synchronised 
behaviour of neuronal networks \cite{sesy-ref5,sesy-ref6}. Moreover patterns
of synchrony and phase shifts in animal locomotion is gaining importance as a
field of active study \cite{sesy-ref7,sesy-ref8,sesy-ref9,sesy-ref10,%
sesy-ref11,sesy-ref12}. In general, the synchronised networks for analysing or
modelling all these physical or biological situations are constructed by 
coupling basic dynamic units with a well defined connection topology that can 
be nearest neighbour, small world, random or hierarchical architectures. In
addition, in specific applications like communication or neural networks, a
realistic modelling may require the introduction of connection delays due to
finite information transmission or processing speeds among the units
\cite{sesy-ref13}. In any case, it is found that the collective dynamics 
depends crucially on the connection topology \cite{sesy-ref14}.

The simplest yet the most widely used topology in this context is the linear 
array and its combinations. The study of synchronisation in arrays of systems 
was first applied to laser systems \cite{sesy-ref15,sesy-ref16} which has
relevance in optical communication systems. Since then the occurrence of 
synchronisation in coupled map lattices has been extensively studied with many
consequent applications \cite{sesy-ref17}.  
Such systems, with synchronisation in temporally
chaotic but spatially ordered units forming an array, is applied in many 
situations like data driven parallel computing \cite{sesy-ref19}.
However most of these cases studied so far involve continuous systems of
chaotic oscillators.

In this paper, we consider two such regular arrays, one vertical and the other
horizontal, that works under the drive response mechanism, where the 
connection is unidirectional. We find that the former setup leads to 
simultaneous synchronisation while the latter results in sequential 
synchronisation. Here we would like to comment that in most of the connected
networks, the synchronisation is found to occur simultaneously. However the
topology in the linear horizontal array introduced here develops 
synchronisation sequentially and the delay time from one unit to the next can 
be adjusted by external control. This mechanism therefore would be useful for
many technological applications. These two types of synchronisations are
characterised using response time (which is the time for synchronisations to
stabilise), size effect, bunching effect etc. These two arrays can be  further
worked together to produce square lattice networks with desirable or
useful inter connections.

The array is realised here with a two dimensional discrete system or map as the
local unit and a connection that involves a non linear function forming part
of the map function. The stability of the simultaneously synchronised state for
the vertical array is studied by computing the Maximum Conditional
Lyapunov Exponent (MCLE) \cite{sesy-ref20}, so that the minimum coupling 
coefficient required for onset of synchronisation can be deduced. The dependence
of the characteristic response time $\tau_s$ on the coupling coefficient 
$\epsilon$
is 
analysed numerically. A horizontal array with the same dynamics is 
constructed with each unit driven by the previous one, modelling an open flow
system and leading to sequential synchronisation. In this case the time taken 
for the last unit to synchronise is taken as the total response time 
$\tau_s$. The behaviour of its average for different initial conditions and 
size $N$ of the system are studied. The additional time or delay time $\tau_l$
required for the last unit to synchronise after its previous one has 
synchronised is found to saturate with system size. Moreover we note an 
interesting bunching effect where the total $\tau_s$ can be controlled by
varying the value of $\epsilon$ in bunch of $m$ units.

In Section 2, we introduce the basic unit which serves as the driving as well
as the driven systems with identical individual dynamics. The concept of
generalised synchronisation and its stability in the context of 
unidirectionally coupled systems is also discussed. The construction and the 
collective dynamics  of the vertical array and the characterisation of 
simultaneous synchronisation is given in section 3. In section 4, we 
introduce sequential synchronisation and its control due to the bunching effect
of the unidirectionally coupled units. Our concluding remarks are given in 
section 5.

\section{Basic Dynamical unit and Generalised Synchronisation}
The basic unit used here for the present analysis of synchronisation in arrays
is a two dimensional discrete systems, which serves both as the driving and 
driven systems defined in the phase space $\overline{X}(n)=
\left(X(n), Y(n)\right)$. The specific system chosen for this work is the 
Gumowski-Mira recurrence relation \cite{sesy-ref21} given as 
\begin{align}
X(n+1) &= Y(n)+a\left(1-b Y{(n)}^2\right) Y{(n)}+ f(X(n)).\notag\\
Y(n+1) &=-X(n)+f(X(n+1)).\label{sesy-eq2.1}
\end{align}
where $f(X(n))=\mu X(n)+\dfrac{2(1-\mu) X^2(n)}{1+X^2(n)}$ and\\
$n$ refers to the discrete time index.

Our earlier investigations in this system reveal that \eqref{sesy-eq2.1}
is capable of giving rise to many interesting two dimensional patterns in 
$(X,Y)$ plane that depend very sensitively on the control parameter $\mu$
\cite{sesy-ref22}. This can be exploited in decision making algorithms and 
control techniques for computing and communications. We have tried three
different coupling schemes in two such systems \cite{sesy-ref23} and found
that they are capable of total or lag synchronisation in periodic,
quasi periodic or chaotic states, when $N$ such systems are geometrically set 
to form a vertical or horizontal array and driven unidirectionally, they are
capable of synchronising to the same chaotic state.

In the context of unidirectionally coupled systems, the type of synchronised
behaviour called generalised synchronisation has been attracting much attention
recently \cite{sesy-ref24,sesy-ref25}. Here the states of the driving system
$\overline{X}_d$ and the driven system $\overline{X}_{dr}$ are dynamically
related by a function $F$ such that the relation $\overline{X}_{dr}(t)=
F(\overline{X}_d(t))$ is true once the transients are over. The form of
$F$ can be smooth or fractal and in either case, the procedure for finding the 
same can be complicated. Hence often an auxiliary system identical to the 
driven system is introduced as $X_a(t)$. The initial conditions of 
$X_{dr}$ and $X_a$ are taken different (both being individually chaotic in 
dynamics) but lying in the basin of the same attractor. Once the 
transients have settled, the dynamical equivalence of ${X}_{dr}(t)$ and 
$X_a(t)$ is taken as an indication of generalised synchronisation between
$X_d(t)$ and $X_{dr}(t)$.

\section{Simultaneous Synchronisation in a Vertical Array}
\begin{figure}[h]
\centerline{\epsfig{file=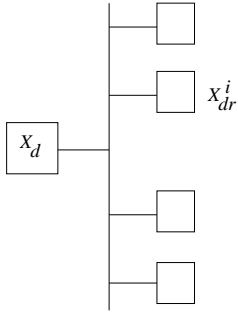, width=.35\linewidth}}
\caption{Schematic view of the construction of a vertical 
array}\label{sesy-fig1}
\end{figure}

We extend the above concept to construct a vertical array of $N$ identical
systems,
$\left[ X_{dr}^1(n), X_{dr}^2(n) \cdots X_{dr}^N(n)\right]$; each driven
independently by $X_d(n)$. All the systems are identical and individually
evolve according to \eqref{sesy-eq2.1}. Fig.~\ref{sesy-fig1} show the above
scheme of construction of vertical arrays.

Here the driving system follows the dynamics
\begin{align}
X_d(n+1) &= Y_d(n)+a\left(1-b Y_d^2{(n)}\right) Y_d{(n)}+ f(X_d(n)).\notag\\
Y_d(n+1) &=-X_d(n)+f(X_d(n+1)).\label{sesy-eq3.1}
\end{align}
with $f(X_d(n))=\mu_d X_d(n)+\dfrac{2(1-\mu) X_d^2(n)}{1+X_d^2(n)}$.\\
The $i^{\text{th}}$ driven unit in the vertical array has the dynamics
\begin{align}
X_{dr}^i(n+1) &= 
Y_{dr}^i(n)+a\left(1-b Y_{dr}^{i^2}{(n)}\right) Y_{dr}^i{(n)}\notag\\
&\quad+ f(X_{dr}^i(n))+\epsilon \left(f(X_d(n))
-f(X_{dr}^i(n))\right)\notag\\
Y_{dr}^i(n+1) &=-X_{dr}^i(n)+f(X_{dr}^i(n+1)).\label{sesy-eq3.2}
\end{align}
with $f(X_{dr}^i(n))=\mu_{dr} X_{dr}^i(n)+\dfrac{2(1-\mu_{dr}) 
X_{dr}^{i^2}(n)}{1+X_{dr}^{i^2}(n)}$ where $\epsilon$ is the coupling coefficient of the 
unidirectional coupling applied to the $X$ variable through the function 
$f(x)$. The parameters $a$ and $b$ are set as $a=0.008$ and $b=0.05$. The 
total number of units considered is $N=51$. The value of $\mu_{dr}$ is chosen
to be the same for all the $50$ driven units. We can realise synchronisation
for different combinations of values of $\mu_{dr}$ and $\mu_d$ with $\mu_d$, in
general different from $\mu_{dr}$. For the special case of $\mu_d=\mu_{dr}$
all the $51$ units synchronise including the driving system, when started with
different initial conditions.

\begin{figure}[h]
\centerline{\epsfig{file=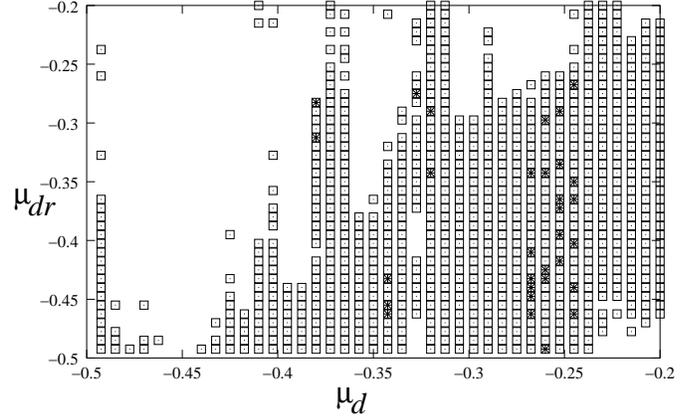, width=\linewidth}}
\caption{Points in the $\mu_d-\mu_{dr}$ plane for which synchronisation
of the driven systems are possible with $\epsilon=0.9$. Points marked 
$\divideontimes$
indicates $\mu_d$, $\mu_{dr}$ values leading to synchronised states with
periodicity less than 15. Points marked $\boxdot$ correspond to 
synchronisation in higher periodic states or mostly chaotic 
states.}\label{sesy-fig2}
\end{figure}

Fixing the value of coupling coefficient $\epsilon=0.9$, the values of $\mu_d$,
$\mu_{dr}$ for which synchronisation is feasible in the $50$ driven systems
is plotted in fig.~\ref{sesy-fig2}. 
In the parameter plane considered here
in the range $-0.2<\mu_d<-0.5$, $-0.2<\mu_{dr}<-0.5$, the points marked $\divideontimes$
indicates $(\mu_d,\mu_{dr})$ values leading to synchronised periodic state
with periodicity less than 15. Points marked $\boxdot$ indicates 
synchronisation
in higher periodic state or mostly chaotic states.

\begin{figure}[h]
\subfigure[]{\epsfig{file=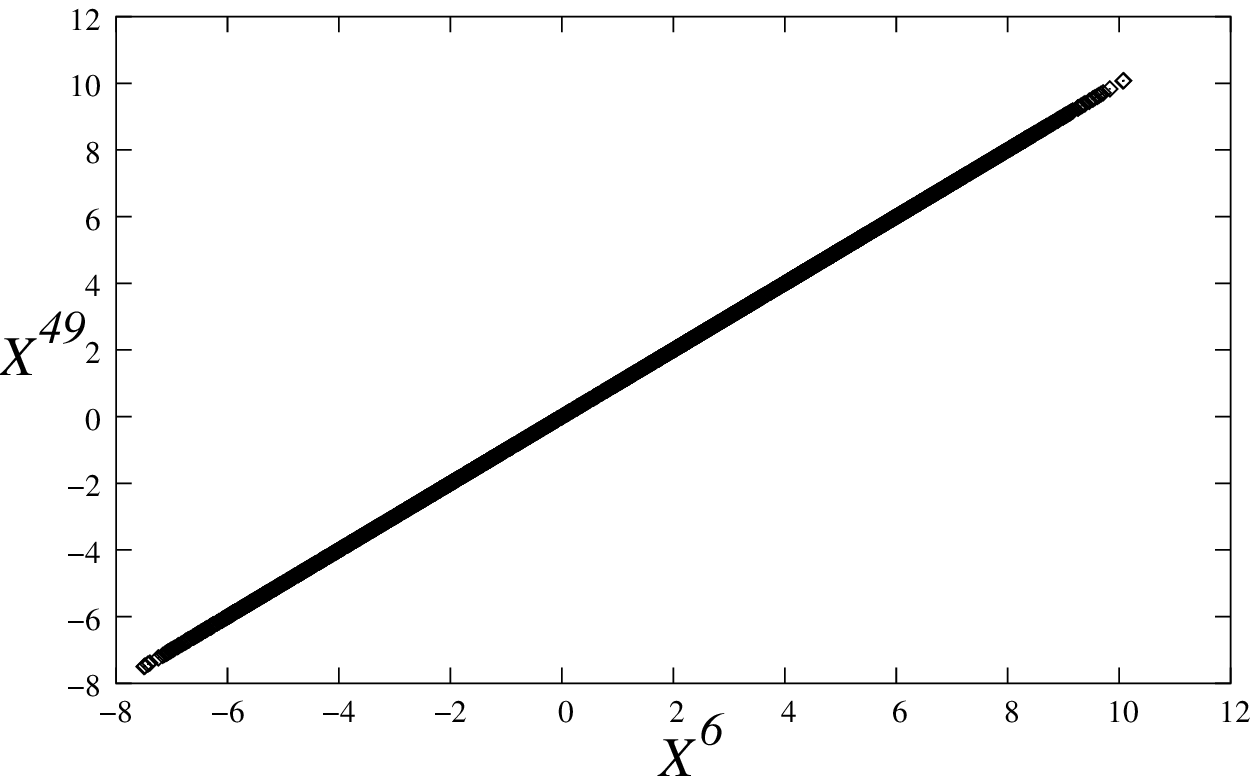, width=\linewidth}}\\
\subfigure[]{\epsfig{file=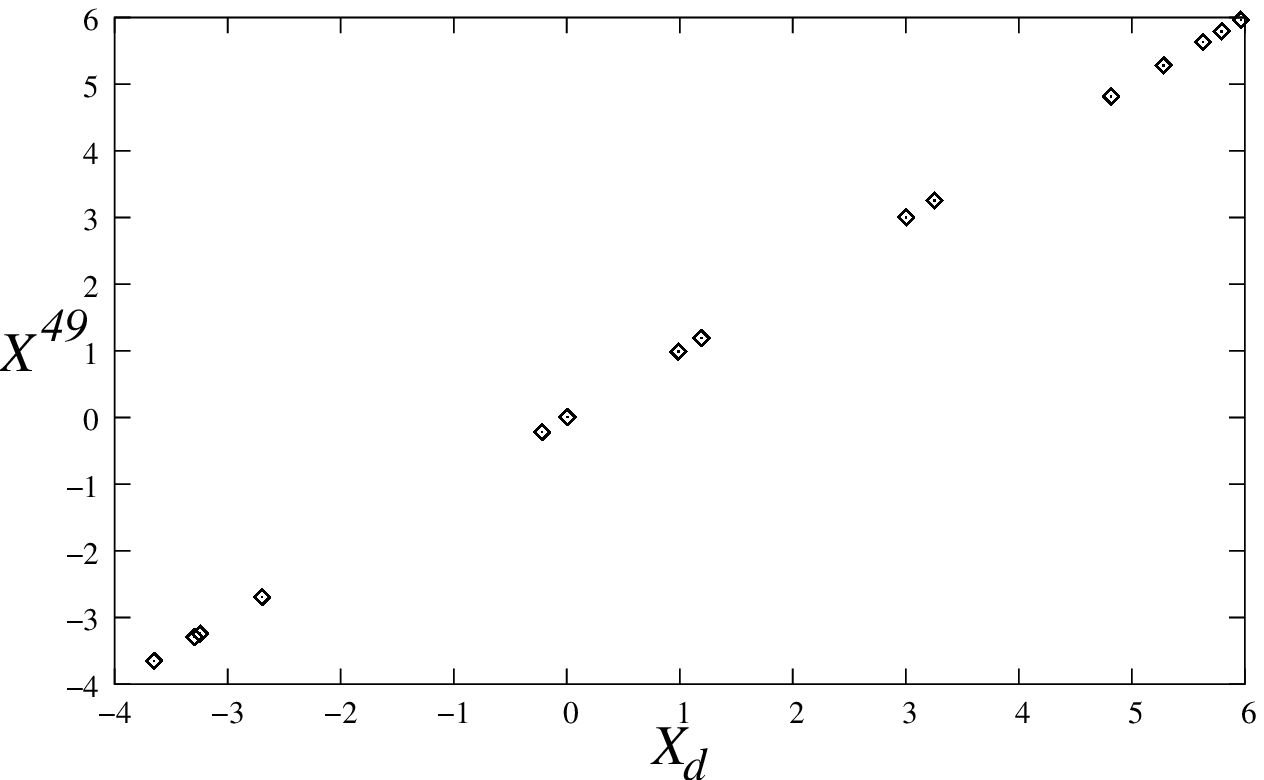, width=\linewidth}}
\caption{(a) Chaotic synchronisation between two participating driven
systems with $\epsilon=1.6$, $\mu_d=\mu_{dr}=-0.39$. Here the iterates of
the $X$ variable of the $6^{\text{th}}$ and $49^{\text{th}}$ units are
plotted. (b) Synchronised periodic 15 cycle for 
$\mu_d=\mu_{dr}=-0.23$ with $\epsilon=1.6$. The iterates of the $X$ variable
of the driving system and $49^{\text{th}}$ driven unit is 
plotted.}\label{sesy-fig3}
\end{figure}

For specific cases like $\mu_d=\mu_{dr}=-0.39$ both the driving system and 
driven systems are in chaotic state individually. With $\epsilon=1.56$ all the
$50$ driven systems are synchronised in the chaotic state while the 
driving system is asynchronous with them. However when $\epsilon$ is slightly
increased to $1.6$ all the $51$ units are found to synchronise in the 
chaotic state. Fig.~\ref{sesy-fig3}a gives this chaotic synchronisation 
between  two participating driven systems for $\epsilon=1.6$, where the 
iterates of the $X$ variable of the $6^{\text{th}}$ and $49^{\text{th}}$ 
units are plotted, after the transients have died out.

For $\mu_d=\mu_{dr}=-0.23$, individually the systems are chaotic. For
$\epsilon=0.9$, all the $N$ driven systems are synchronised to the same 
periodic state of periodicity 15. But the driving system is also synchronised
only when $\epsilon$ is increased to 1.6. Fig.~\ref{sesy-fig3}b gives this
synchronised periodic 15 cycle for $\epsilon=1.6$, where the iterates
of the driving system and the $49^{\text{th}}$ unit are plotted. Thus the 
driving system and the driven system can be simultaneously synchronised only
when $\mu_d=\mu_{dr}$ and when $\epsilon$ is very large ie 
$\epsilon\sim 1.6$.

\begin{figure}[h]
\centerline{\epsfig{file=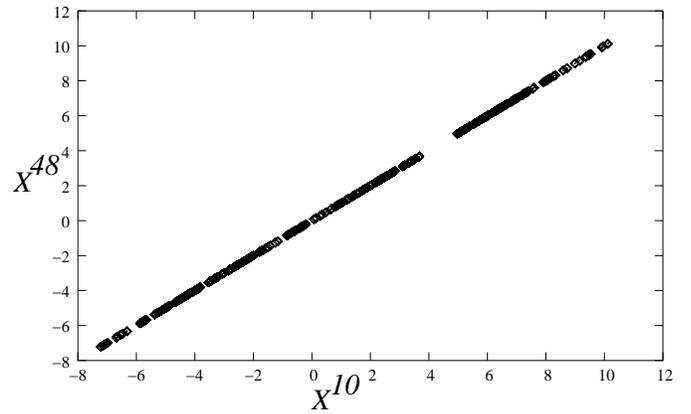, width=\linewidth}}
\caption{Synchronised chaotic states between two driven systems for
$\mu_d=-0.2$, $\mu_{dr}=-0.39$ with $\epsilon=0.9$. Here the iterates of the
$X$ variable of the $48^{\text{th}}$ unit and the $10^{\text{th}}$ units are
plotted.}\label{sesy-fig4}
\end{figure}

For $\mu_d=-0.2$, the driving system is in periodic 8 cycle. For 
$\mu_{dr}=-0.39$ the driven systems are individually chaotic. For
$\epsilon=0.9$ all the $N$ driven systems are synchronised in the chaotic
state. Fig.~\ref{sesy-fig4} shows the synchronised chaotic state for the above 
case. Here the iterates of the $X$ variable of the $48^{\text{th}}$ unit and
$10^{\text{th}}$ unit are plotted after the transients have died out.

The condition for the stability of generalised synchronisation is discussed
using the Maximal Conditional Lyapunov Exponent $\lambda_{MCLE}$ 
\cite{sesy-ref26,sesy-ref27}. Here the Lyapunov Exponent of the driven system
is calculated and it is different from the uncoupled system, since it
depends on the dynamics of the driving system also. The condition for the 
stability of generalised synchronisation is that $\lambda_{MCLE}$ should
be negative \cite{sesy-ref28}. From equations \eqref{sesy-eq3.1} and
\eqref{sesy-eq3.2} the Jacobian matrix for the $i^{\text{th}}$ unit can be
written as
\begin{equation}\label{sesy-eq3.3}
M=\begin{bmatrix}
(1-\epsilon)\frac{\partial F^i}{\partial X^i} & 
\frac{\partial F^i}{\partial Y^i}\\
\frac{\partial G^i}{\partial X^i} & \frac{\partial G^i}{\partial Y^i}
\end{bmatrix}
\end{equation}
where $X^i_{dr}(n+1)=F^i(X,Y)$ and $Y^i_{dr}(n+1)=G^i(X,Y)$.\\
If $\sigma^1$ and $\sigma^2$ are the eigen values of the product of the 
Jacobian matrices at every iteration such that $\sigma^1>\sigma^2$,
then \cite{sesy-ref29}
\begin{equation}\label{sesy-eq3.4}
\lambda_{MCLE}=\lim_{m\rightarrow \infty}\frac{1}{m} ln |\sigma^1|
\end{equation}
$\lambda_{MCLE}$ can be calculated numerically for different $\epsilon$ values
using \eqref{sesy-eq3.4}.

\begin{figure}[h]
\centerline{\epsfig{file=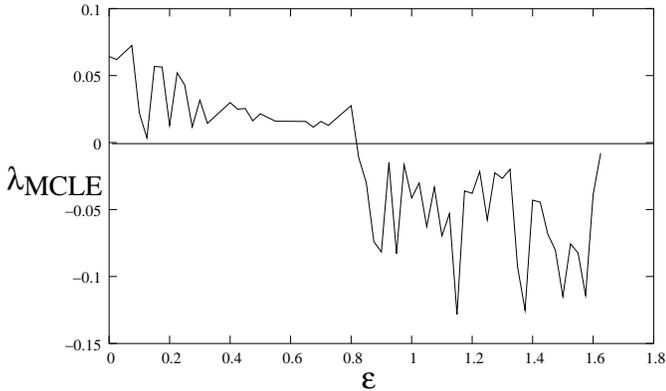, width=\linewidth}}
\caption{The variation of Maximum Conditional Lyapunov Exponent 
($\lambda_{\text{MCLE}}$) 
of the driven system with the coupling coefficient $\epsilon$. 
Here $\mu_d=-0.2$,
$\mu_{dr}=-0.39$. The minimum coupling coefficient for synchronisation
$\epsilon_{\min}=0.829$.}\label{sesy-fig5}
\end{figure}

We consider the case $\mu_d=-0.2$ and $\mu_{dr}=-0.39$ which we have
discussed above and calculate $\lambda_{MCLE}$ for different $\epsilon$
values. Calculations are done for 10000 iterates after leaving initial
70000 iterates as transients. In fig.~\ref{sesy-fig5} 
the values of $\lambda_{MCLE}$ for different
$\epsilon$ values are plotted. It is found that, $\lambda_{MCLE}$ 
crosses zero at $\epsilon=0.829$, which is the minimum value of $\epsilon$
\cite{sesy-ref20} viz $\epsilon_{\min}$ such that for 
$\epsilon>\epsilon_{\min}$ the synchronised state is stable.

\begin{figure}[h]
\centerline{\epsfig{file=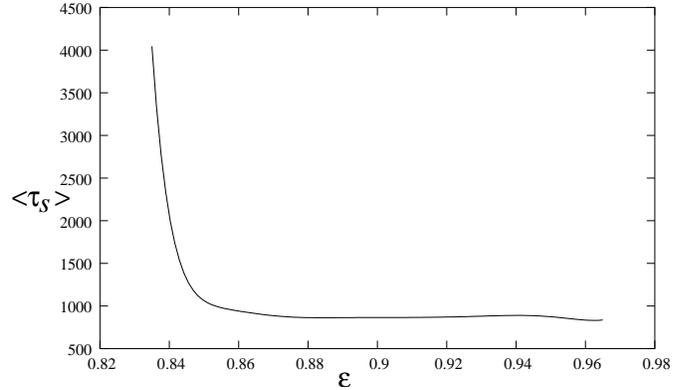, width=\linewidth}}
\caption{The variation of the response time $\langle \tau_s\rangle$ which is
the total time for synchronisation with coupling coefficient $\epsilon$ for
the $50^{\text{th}}$ unit. $\langle \tau_s\rangle$ is almost constant for
values of $\epsilon > 0.87$.}\label{sesy-fig6}
\end{figure}

For the above case the coupling coefficient is varied in steps from $0.84$ to
$0.97$ and the time taken for reaching synchronisation in the driven systems 
is noted. Fig.~\ref{sesy-fig6} gives the variation of thus average response
time $\tau_s$ (averaged over 
10 different initial conditions) with $\epsilon$ for the
$50^{\text{th}}$ unit. It is interesting to note that $\langle \tau_s\rangle$ 
is almost constant for values of $\epsilon>0.87$. In this case since 
the synchronisation
is simultaneous and coupling is unidirectional and similar, the average
$\langle \tau_s \rangle$ is independent of the size of the array $N$.

\section{Sequential synchronisation in a Horizontal Array}
\begin{figure}[h]
\centerline{\epsfig{file=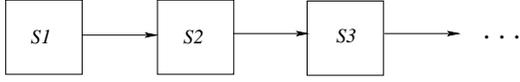, width=.8\linewidth}}
\caption{Schematic view of unidirectional coupling in a flow 
system.}\label{sesy-fig7}
\end{figure}

In this section a horizontal  array of $N$ identical systems with open ends,
where each unit is driven by the previous one is introduced. The coupling is
through the nonlinear function $f(X,Y)$ as in the previous case.
Fig.~\ref{sesy-fig7} gives the schematic view of unidirectional coupling in a 
flow which consists of $N$ units.

The $i^{\text{th}}$ unit in the horizontal array follows the dynamics

\noindent\begin{align}
X^i(n+1) &=
Y^i(n)+a\left(1-b Y^{i^2}{(n)}\right) Y^i{(n)}\notag\\
&\quad+f(X^i(n))+\epsilon \left(f(X^{i-1}(n))
-f(X^i(n))\right)\notag\\
Y^i(n+1) &=-X^i(n)+f(X^i(n+1)).\label{sesy-eq4.1}
\end{align}
with $f(X^i(n))=\mu X^i(n)+\dfrac{2(1-\mu) X^{i^2}{(n)}}{1+X^{i^2}{(n)}}$.

The control parameter $\mu$ in the same for all the units such that the units 
are chaotic individually. This set up is found to give rise to sequential 
synchronisation in the array. In our calculations we consider an array of 51
units. This can be extended to any number of units $N$.

\begin{figure}[h]
\centerline{\epsfig{file=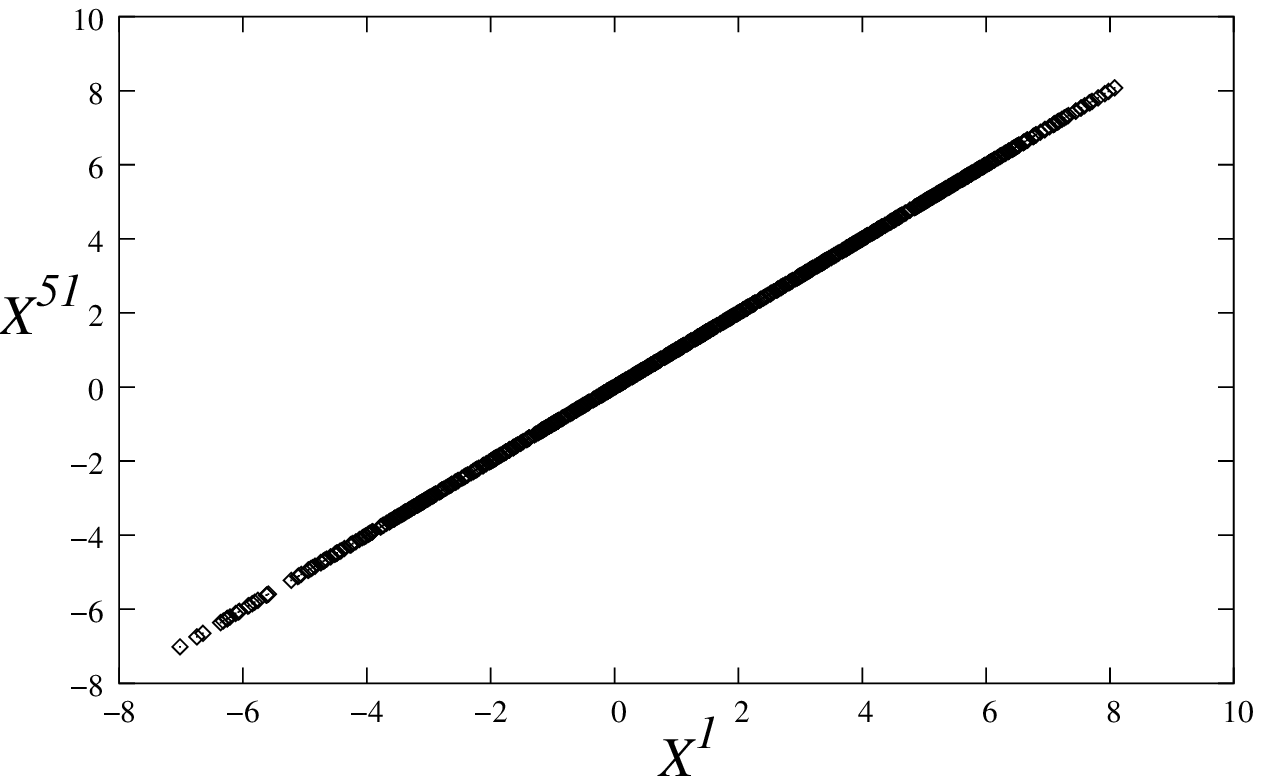, width=\linewidth}}
\caption{Synchronised chaotic state for $\mu=-0.23$ with $\epsilon=1.9$.
The iterates of the $X$ variable of the first unit and the last unit
($51^{\text{st}}$ unit) are plotted.}\label{sesy-fig8}
\end{figure}

\begin{figure}[h]
\centerline{\epsfig{file=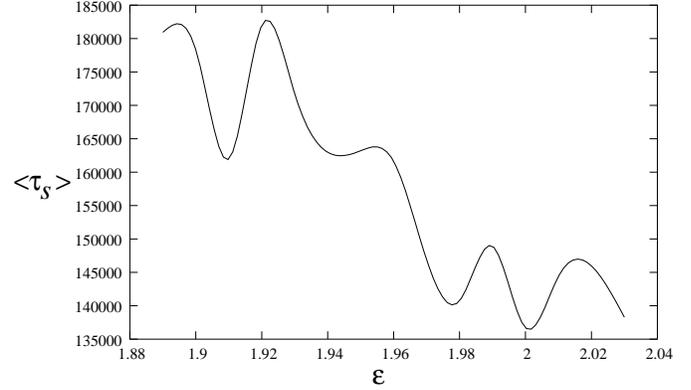, width=\linewidth}}
\caption{The variation of total response time $\langle \tau_s\rangle$ of the 
$51^{\text{st}}$ unit with coupling coefficient $\epsilon$.
$\langle \tau_s\rangle$ has a minimum value for $\epsilon=2$.}\label{sesy-fig9}
\end{figure}

For $\mu=-0.23$ where the individual systems are chaotic, and coupling coefficient
$\epsilon=1.9$, we find that synchronisation sets in sequentially with the
$2^{\text{nd}}$ synchronising after the first, the third after the second and
so on. The time taken by the last unit to synchronise is taken as 
$\langle \tau_s \rangle$
which is the average total response time for the whole array. 
Fig.~\ref{sesy-fig8}
shows this synchronised chaotic state after the last unit has synchronised. The
 $\langle \tau_s\rangle$ is found to vary with coupling coefficient $\epsilon$
as shown in Fig.~\ref{sesy-fig9}. 
It is found that $\langle\tau_s\rangle$ has a minimum value
for a particular $\epsilon$ which in this case is $\epsilon=2$.

\begin{figure}[h]
\centerline{\epsfig{file=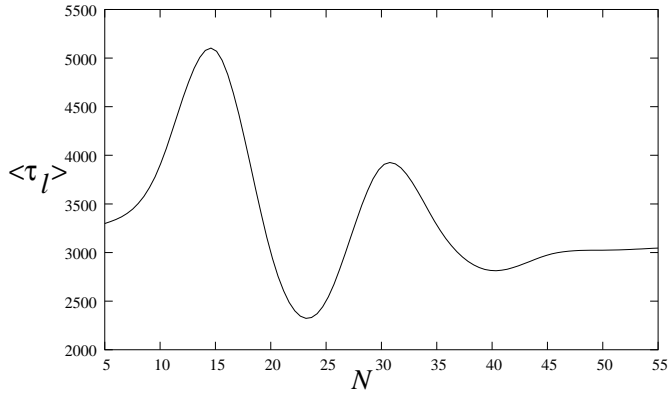, width=\linewidth}}
\caption{The delay time $\tau_l$ which is the additional time taken for the 
$N^{\text{th}}$ unit to synchronise after its previous one has 
synchronised is found to saturate with system size $N$. Beyond $N=35$,
$\tau_l$ is almost a constant.}\label{sesy-fig10}
\end{figure}

The delay time $\tau_l$ ie, the additional taken for the $N^{\text{th}}$ unit
to synchronise after its previous one has synchronised is defined as 
$\tau_l=\tau^N_s-\tau^{N-1}_s$. This $\tau_l$ is found to saturate with the 
system size \cite{sesy-ref30} as shown in fig.~\ref{sesy-fig10}. Beyond
$N=35$, $\tau_l$ is almost constant.

An interesting observation in this horizontal array of units is a bunching
effect that reflects in the total response time $\langle\tau_s\rangle$. 
For this 
instead of fixing the same value for the coupling coefficient $\epsilon$ for all
the units, we fix its value for a particular number of units and increase
it in steps for the next bunch and so on. Then the total $\langle\tau_s\rangle$
is found to 
be smaller compared to the previous case of the same $\epsilon$ for all the 
units. Moreover this time depends on the size of the bunch and is minimum for
a certain number of units in each bunch.

\begin{figure}[h]
\begin{center}
\mbox{
\epsfig{file=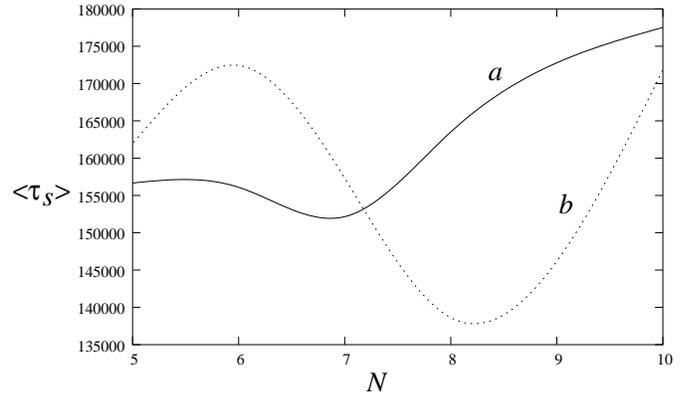, width=\linewidth}}
\end{center}
\caption{Change in the response time $\langle \tau_s\rangle$ of the 
last unit with bunch size $m$. Here $\mu=-0.23$. \textit{Curve a} is for 
$\epsilon_{\max}=1.91$, when the bunch size $m=7$, the response time
$\langle \tau_s\rangle$ is minimum. \textit{b} is for $\epsilon_{\max}=2.01$, 
the response time $\langle \tau_s\rangle$ is minimum for $m=8$. In both cases 
$\langle \tau_s\rangle$ is found to be less than the case when we apply 
$\epsilon_{\max}$ to all the units.}\label{sesy-fig11}
\end{figure}

We report a few specific cases. With $\mu=-0.23$ the value of $\epsilon$
is increased in steps of $0.001$ for each bunch so that $\epsilon$ for the 
last bunch is $\epsilon_{\max}=2.01$ for different bunch sizes. In each case
the total response time $\langle \tau_s\rangle$ is found. 
Fig. \ref{sesy-fig11} shows how the response time 
$\langle \tau_s\rangle$ changes
with the variation in the bunch size $m$, ie., number of units in each bunch.
The response time $\langle \tau_s\rangle$ is minimum when the bunch size
$m=8$. For $m=8$, $\langle \tau_s\rangle=138612$ iterations, whereas when
$\epsilon=\epsilon_{\max}=2.01$ for all the units, the response time 
$\langle \tau_s\rangle=144404$ iterations.

\begin{figure}[h]
\begin{center}
\centerline{\epsfig{file=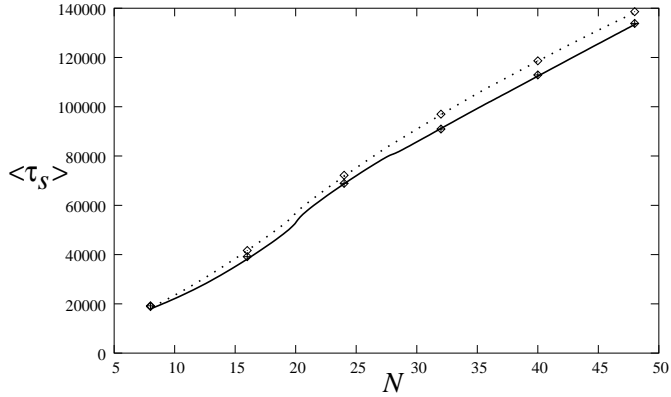, width=\linewidth}}
\caption{The response time $\langle \tau_s\rangle$ is plotted for different
units for $\mu=-0.23$, $\epsilon_{\max}=2.01$.
The dotted line gives the total response time $\langle \tau_s\rangle$ of 
8$^{\text{th}}$ unit, 16$^{\text{th}}$ unit, 24$^{\text{th}}$ unit etc.
when $\epsilon_{\max}=2.01$ is applied to all the units.
Full line gives the total response time $\langle \tau_s\rangle$ 
of 8$^{\text{th}}$ unit
(last unit of 1$^{\text{st}}$ bunch), 16$^{\text{th}}$ unit (last unit of 
2$^{\text{nd}}$ bunch), 24$^{\text{th}}$ unit (last  unit of 3$^{\text{rd}}$
bunch) etc. when $\epsilon$ is increased in steps for each bunch so that
$\epsilon_{\max}=2.01$ for the last bunch. Here $2.005\leq \epsilon\leq
2.01$ with step size 0.001. Thus bunching can control the total
response time $\langle \tau_s\rangle$ of the array.}\label{sesy-fig12}
\end{center}
\end{figure}

As a second case for same $\mu=-0.23 \epsilon_{\max}$  is taken as $1.91$ and
calculations repeated as above. In this case $\langle \tau_s\rangle$ is found
to be minimum and is $152150$ iterations when the size of the bunch is 
$m=7$ as shown in Fig. \ref{sesy-fig11}. If $\epsilon=1.91$ for all the units,
$\langle \tau_s\rangle$ is $161953$ iterations. We observe that the decrease in
$\langle \tau_s\rangle$ for the whole array due to bunching must be 
reflected in the response time of each bunch. So for the minimum case, the 
response times for the last unit of the first bunch 
(ie., 8$^{\text{th}}$ unit), last unit of the second bunch 
(ie., 16$^{\text{th}}$ unit) etc. are noted with bunching. The same quantity
with $\epsilon$ same for all the units ie., without bunching are also
noted. $\langle \tau_s\rangle$ thus obtained are plotted against the respective
units in Fig. \ref{sesy-fig12}. 
It is found that except for the 8$^{\text{th}}$ unit in the first bunch 
the response time is less in the case of bunching.

\section{Conclusion}
In this work we report how synchronisation in an array of systems can be made
more efficient and  flexible to  suit specific applications. We consider
two such arrays, vertical and horizontal, working under the drive-response
mechanism with a two-dimensional discrete systems as the unit dynamics.

It is observed that synchronisation sets in all the systems
simultaneously in the vertical setup. The minimum value of the coupling 
coefficient $\epsilon$ required for stability of synchronisation is computed 
numerically from the Maximum Conditional Lyapunov Exponent. The possible 
choices of parameters for stable states of synchrony are isolated. 
The specific cases of chaotic single units 
synchronising to periodic and chaotic synchronised states and periodic single
units stabilising to chaotic states of synchronisation are considered 
in detail. The average response time required to overcome the transients is 
found to saturate beyond a certain value of $\epsilon$. The horizontal array
exhibits many interesting features useful for technological applications.
In this case the synchronisation sets in sequentially from unit to unit along
the array since the coupling is unidirectional. The total response time for 
the whole array has a minimum as $\epsilon$ is varied. The additional time
required for the last unit to synchronise after the previous one is found to
saturate with system size.

We further note that the total response time for the whole array can be 
reduced by introducing bunching with step wise increase of $\epsilon$ from 
bunch to bunch. There exists a specific bunch size giving minimum time which
depends on the choice of the parameter and the maximum $\epsilon$ given to
the last bunch. This makes the sequential synchronisation flexible and 
controllable to suit specific applications. 

At present we do not find any specific reasons for the above findings to 
depend on the unit chosen. For different choices of the unit dynamics the 
behaviour should be qualitatively similar. To establish this generality and
applicability of the present technique, we are trying it out for a number 
of systems.
The results will be published elsewhere.

\medskip\noindent\textbf{Acknowledgement}\\
K. A. thanks University Grants Commission, New Delhi for deputation
under Faculty Improvement Programme. G. A. thanks IUCAA, Pune for hospitality 
and computer facility.

\end{document}